# Tracking ultrafast solid-state dynamics using high harmonic spectroscopy

**Authors:** Mina R. Bionta[1,†,*], Elissa Haddad[1], Adrien Leblanc[1], Vincent Gruson[1,2], Philippe Lassonde[1], Heide Ibrahim[1], Jérémie Chaillou[1], Nicolas Émond[1], Martin R. Otto[3], Bradley J. Siwick[3], Mohamed Chaker[1], and François Légaré[1,*]

**Affiliations:**

[1]Centre Énergie Matériaux et Télécommunications, Institut National de la Recherche Scientifique, 1650 Boulevard Lionel-Boulet, Varennes, Québec, J3X 1S2, Canada.

[2]Department of Physics, The Ohio State University, 191 West Woodruff Avenue, Columbus, Ohio, 43210, USA.

[3]Department of Physics and Department of Chemistry, Center for the Physics of Materials, McGill University, 801 Sherbrooke Street W, Montreal, Québec, H3A 2K6, Canada.

[†]Present address: Research Laboratory of Electronics, Massachusetts Institute of Technology, 77 Massachusetts Avenue, Cambridge, Massachusetts, 02139, USA.

[*]Correspondence to: mbionta@mit.edu; legare@emt.inrs.ca

**Abstract:**

We establish time-resolved high harmonic generation (tr-HHG) as a powerful spectroscopy for photoinduced dynamics in strongly correlated materials through a detailed investigation of the insulator-to-metal transitions in vanadium dioxide. We benchmark our technique by comparing our measurements to established momentum-resolved ultrafast electron diffraction, and theoretical density functional calculations. Tr-HHG allows distinguishing of individual dynamic channels, including a transition to a thermodynamically hidden phase. In addition, the HHG yield is shown to be modulated at a frequency characteristic of a coherent phonon in the equilibrium monoclinic phase over a wide range of excitation fluences. These results demonstrate that tr-HHG is capable of tracking complex dynamics in solids through its sensitivity to the band structure.

**Introduction:**

As ultrashort light and electron sources have become more advanced, real-time dynamics of complex states of matter may be understood in greater detail[1,2]. High-harmonic generation (HHG) in atoms and molecules is a well-known process for generating table-top ultrafast sources from the extreme ultraviolet (EUV) to the soft X-ray spectral range, for probing dynamics in matter[1,3,4]. High-harmonic spectroscopy[3] has had great success probing electronic structure and dynamics in atoms and molecules[5–7], as well as in following chemical reactions in the gas phase[8–10]. Extending high-harmonic generation (HHG) to solid state systems[11–13] allows for highly sensitive probing of band structure[14–18].

HHG was recently extended to solids, as first reported by Ghimire *et al.*, who observed high harmonics generated from ZnO[11]. Over the years, two types of harmonics generated in solids have been identified: intraband and interband[12–14,17–21]. In both cases, the first step consists of multiphoton/tunneling excitation from a valence to a conduction band. For intraband harmonics, the generated radiation originates from a nonlinear band current driven by the laser field; for interband harmonics, the generation mechanism involves recombination of the accelerated electron and hole similar to HHG from atoms and molecules. In general, harmonics with photon energies below the bandgap are intraband in nature, while those that are more energetic are interband[17,20]. In both cases, the harmonic spectra generated by solids should be a very sensitive probe of ultrafast phenomena that involve changes to the band structure near the Fermi level. Thus far, most HHG measurements in solids have been performed on static systems; however, pump-probe experiments on ZnO have revealed that the intraband HHG yield is suppressed upon photodoping and remains quenched for several tens of picoseconds until the electrons return to the ground state[14]. The theory of HHG in solids is at an early stage, but undergoing rapid development and suggest that several of the characteristic features of Mott-Hubbard physics should impact directly the HHG yield from strongly correlated materials[22,23]. This picture of HHG suggests a sensitivity to a range of nonequilibrium phenomena occurring in strongly correlated materials, since many of these – including insulator-to-metal transitions – are associated with a redistribution of spectral weight/optical conductivity over several eV, the energy scale on which HHG is most sensitive.

The current study takes an experimental approach to test this hypothesis by investigating how the time-resolved HHG yield changes during the photoinduced insulator-to-metal phase transitions in the strongly correlated material vanadium dioxide ($VO_2$). Strongly correlated materials are of particular interested because any optical excitation that modifies the interplay between lattice, charge, orbital, or spin degrees of freedom can result in dramatic transformations in their properties[24–30]. In practice, the optical perturbation is typically associated with the excitation of carriers (photodoping) or strongly driving a specific infrared-active lattice mode (phonon-pumping). Generally, a complete picture of these transient, optically-induced phenomena in materials requires the use of complementary techniques to reveal both the structure and properties of the nonequilibrium states, and the associated changes to the coupling between various degrees of freedom. Time-resolved techniques including X-ray or electron scattering[2,31], photoelectron, and transient spectroscopies ranging from the terahertz to the X-ray regime have become the methods of choice for this burgeoning field[32,33].

Photoexcited $VO_2$ exhibits a rich phenomenology enabled by its multi-band Mott-Hubbard character[32–39]. Recently, it has been shown that there are two qualitatively distinct photoinduced insulator-to-metal transitions in VO$_2$ following photoexcitation[32,33,37]. The first one, accessible at relatively high pump fluence, is an analog of the equilibrium phase transition, and is associated with the lattice-structural transition between the monoclinic insulator ($M_1$) and the rutile metallic ($R$) crystallography expected from the equilibrium phase diagram. The second one, accessible at lower pump fluence, has no equilibrium/thermodynamic analog and yields a metastable, monoclinic metal phase ($\mathcal{M}$) that retains the crystallographic symmetry of its parent equilibrium monoclinic phase, but exhibits a novel 1D antiferroelectric charge order not present at equilibrium[33]. Furthermore, a hidden phase transition of primarily electronic character has also been identified through numerical simulations using density functional theory[34].

In Fig. 1a, schematic partial density of states diagrams for these three phases ($M_1$, $\mathcal{M}$, and $R$) are presented. There is broad agreement that the principal changes in band structure associated with these transitions occur in the bands formed by the V$_{3d}$ states of $t_{2g}$ symmetry ($d_{x2y2} \rightarrow d_\parallel$, $d_{xy} \rightarrow d_\pi^\perp$, $d_{xz} \rightarrow d_\pi^\parallel$)[40]. In the $M_1$ phase, the optical bandgap is formed between the $d_\parallel$ and $d_{\pi*}^\parallel$ bands. In the $R$ phase, the bandgap collapses and the $d$ bands are effectively degenerate at the Fermi level. In the $\mathcal{M}$ phase, recent calculations suggest a re-ordering of $d_\parallel$ and $d_{\pi*}$ bands yielding a partial overlap at the Fermi level[34]. We show that tr-HHG reports on each of these previously demonstrated features in all three phases of VO$_2$, benchmarking the technique as an all-optical method for probing dynamics in strongly correlated materials.

**Results:**

In these experiments, a mid-infrared laser pulse (MIR) called the driver, centered at either 10 μm or 7 μm, with a pulse duration of 80 fs and peak intensity in the range of ~2x10$^{12}$ W cm$^{-2}$, is used to drive HHG from a 100 nm thick, epitaxial VO$_2$ sample[35] as shown in Fig. 2a. The VO$_2$ sample, in the $M_1$ phase at a room temperature of 20 °C, is photoexcited with a 50 fs laser pulse centered at 1.5 µm, called the pump, to initiate the IMT. The generation of the 5$^{th}$ harmonic of the 10 µm HHG driver (at 2.0 µm) or 3$^{rd}$ harmonic of the 7 µm driver (at 2.3 µm) is measured as a function of time delay between the pump and driver for various pump fluences by a spectrometer or photodiode, as presented in Fig. 2a. Given the driver intensity and wavelength, and that the measured harmonic is below the band gap of VO$_2$ (0.68 eV – 1.815 µm), the HHG signal is dominated by intraband harmonics[17,20]. These measurements were taken in both the forward and backward direction (for backward measurements, see Supplementary Information Sec. S3). Furthermore, pump-probe transmissivity measurements are performed with infrared (IR) probe pulses at 1.7 µm.

In Fig. 2b, we present HHG spectra for two temperatures, 293 K (20 °C) and 373 K (100 °C) driven by 10 µm. When heated above ~343 K (70 °C)[36], VO$_2$ undergoes an insulator-to-metal transition that is associated with a change in crystallography from the monoclinic insulator ($M_1$) to metallic rutile ($R$) phase. In the $M_1$ phase, we observe up to the 7$^{th}$ harmonic in the HHG spectra while for the $R$ phase, there is significant suppression of the harmonic yield with no measurable harmonics (Fig. 2b). This is consistent with the flattening of conduction bands in $R$ phase VO$_2$[40,41] (Fig. 1b), decreasing the anharmonicity required for HHG and suppressing the generation of harmonics. The same behavior is observed in the backward direction (see Supplementary Information Sec. S3). We note that the $R$ phase is metallic and are unaware of any observation of HHG from metals. Thus, we expect a significant drop of the harmonic yield in the time-resolved experiments for the photoinduced IMT.

Before we present the tr-HHG measurements, we report the pump-probe transmissivity measurements. In Fig. 3a and b, at pump fluences greater than 6 mJ cm$^{-2}$, we observe a drop in the optical transmissivity at 1.7 μm whose magnitude monotonically increases with the pump fluence. This initial drop, limited by the pump pulse duration[35,37], is followed by a flat response within the temporal window of 10 ps. Within this time scale, there is no recovery of the IR transmissivity. Similar measurements have been performed with a 1.3 μm pump and 4 μm probe showing these same dynamics with no recovery in transmissivity observed.

Fig. 3c and d show the tr-HHG measurements for the 5$^{th}$ harmonic driven by 10 μm. At high pump fluence, the measurements are qualitatively similar to the transmissivity measurements at 1.7 μm. Pump fluences greater than 35 mJ cm$^{-2}$ lead to a nearly complete suppression of the harmonic yield at time zero with no recovery observed over 10 ps (see Fig. 3c). These dynamics are very similar to recent tr-HHG measurements in the semiconductor ZnO where the HHG yield is greatly suppressed by photodoping, recovering only after several tens of picoseconds once the system returns to the ground electronic state[14]. Recent UED measurements[32,33] show that – at these pump fluences – approximately 80% of the film undergoes the photoinduced $M_1$ to $R$ phase transition. Thus, the significant suppression of HHG for VO$_2$ for pump fluences where the material undergoes a transition to the $R$ phase is consistent with the temperature dependent HHG measurements presented previously (see Fig. 2b).

If the pump fluence is very low, ≤ ~3 mJ cm$^{-2}$, both the IR transmissivity and harmonic yield experience a small suppression at time-zero that rapidly recovers (within 500 fs) to very nearly the same signal levels before photoexcitation. This indicates that at low pump fluence no phase transitions are induced, and VO$_2$ rapidly returns to the equilibrium $M_1$ phase (Fig. 3c and d). This observation is also in good agreement with UED and several other time-resolved spectroscopic measurements[32,33,35,38].

In the fluence range from 3-35 mJ cm$^{-2}$, however, there is a striking difference between the time-resolved transmissivity and the tr-HHG measurements. Immediately after photoexcitation, both signals exhibit a drop, limited in time by the pump pulse duration, whose amplitude monotonically increases with pump fluence. This drop is due to the injection of photodoped carriers into the conduction band, inhibiting intraband HHG[14]. However, the tr-HHG yield from VO$_2$ demonstrates a recovery on the picosecond time scale, much faster than the recovery of its own IR transmissivity as well as compared to photodoped ZnO[14]. This recovery is characterized by a bi-exponential with an amplitude that is non-monotonic with pump fluence. This recovery has two timescales; a fast one of ~300 fs followed by a second one of ~1.5 ps (see Fig. 4c, and Supplementary Information Sec. S2). The fast term, here called the $M_1^* \to M_1^{*,b}$ transition, shows reasonable agreement with the recent numerical calculations describing thermalization of the photodoped carrier distributions[34]. The slow term matches UED measurements of the $\mathcal{M}$ phase formation time[32,33]. The behavior is also observed in the backward directed tr-HHG (Supplementary Information Sec. S3) as well as in the tr-HHG dynamics of the 3$^{rd}$ harmonic driven by 7 μm pulses (Fig. 4a and b).

**Discussion:**

From UED measurements[32,33], it is known that the polycrystalline $VO_2$ sample demonstrates a multi-phase heterogeneous response over this range of pump fluences (3-35 mJ cm$^{-2}$) due to the concurrence of the $M_1 \to R$ and $M_1 \to \mathcal{M}$ transitions. UED measurements showed that the phase fraction of the $\mathcal{M}$ phase increases up to a pump fluence of ~20 mJ cm$^{-2}$ and then decreases at higher pump fluence where rutile dominates. This is precisely the non-monotonic behavior of the HHG signal recovery observed by the tr-HHG measurements. The amplitude of the HHG yield recovery is shown in Fig 4d alongside the phase fraction of $\mathcal{M}$ obtained by UED[33].

The tr-HHG results from both driver wavelengths show a non-monotonic recovery in the HHG yield through the pump fluence range studied compared with the IR optical transmissivity, which rapidly decreases and then remains always flat over 10 ps, with no recovery observed. Curiously, the HHG yield from the $\mathcal{M}$ phase appears to be very similar to that of the $M_1$ phase, despite being metallic with a low frequency THz average conductivity ~1/3 that of the equilibrium metallic $R$ phase[33]. This observation seems to be at odds with those shown in Fig. 2b, where the $R$ phase is shown to have negligible HHG yield. The high production of harmonics in the $\mathcal{M}$ phase (see Fig. 3) indicates that despite this metallic character, as highlighted by a drop of transmissivity that persists for tens of picoseconds (see Fig. 3a), the bands near the Fermi level (see Fig. 1b) must maintain the electronic structure required for HHG[12–14,17–21]. This is a salient observation that does not follow from the earlier studies using more conventional optical spectroscopies, nor was it predicted by theory. These observations may be due to the strong correlations in the material and a hopping-like character to the conductivity, or a different sensitivity of optical and HHG yield to electron localization/delocalization processes and shifts in electronic band structure. However, the most likely explanation seems to be that the $\mathcal{M}$ phase is best described as 1D metal with high conductivity limited to a single crystallographic direction. This direction is most likely to be the monoclinic *a*-axis (equivalent to the rutile *c*-axis), along which the vanadium atoms are dimerized and antiferroelectric charge ordering in the $\mathcal{M}$ phase has been observed in UED experiments[33]. If the $\mathcal{M}$ phase is a 1D metal, it could simultaneously present metallic conductivity when probed by tr-THz spectroscopy[33] and maintain a high HHG yield as observed in the current measurements. This suggests that the picture presented in Fig. 1b for the $\mathcal{M}$ phase should be modified to include directions perpendicular and parallel to the monoclinic *a*-axis. Along the *a*-axis, there is band overlap similar to $R$, but perpendicular to the *a*-axis an anharmonic bandgap remains, thus fulfilling the condition for HHG.

Another notable feature of the tr-HHG yield is the presence of coherent oscillations during the recovery (see Fig. 4b). If the IMT is driven impulsively, coherent phonon dynamics may be launched through Raman scattering. These phonon modes, alongside photodoped electrons, provide a pathway for the excess energy from the photoexcitation to reach equilibrium. Raman spectroscopy studies on both the insulating monoclinic phase and rutile metallic phase show distinct active phonon modes: 18 in the monoclinic phase and 4 in the rutile phase[37,42]. The oscillations for low to moderate pump fluences up to 27 mJ cm$^{-2}$ are matched to a period of about 235 fs corresponding to the lowest frequency of the coherent phonon mode of monoclinic VO$_2$ at 4.4 THz[37,42]. The persistence of these phonon dynamics remain for pump fluences much higher than those reported in [37] via time-resolved reflectivity measurements and suggest that the tr-HHG technique can map the coherent phonon dynamics. The origin of this modulation can be either a change in bandgap, thus modulating the photoexcitation probability, or the anharmonicity of the band. At present, our measurements cannot identify the specific mechanism responding to the phonon modes which leads to the modulation of the tr-HHG yield. Yet, our measurement is one of the first instances of tr-HHG of coherent phonon dynamics[43] similar to observation of nuclear dynamics with HHG spectroscopy of photoexcited gas phase molecules[9,10]. The presence of these phonon modes for the monoclinic phase $VO_2$ and not rutile indicates that although there is an initial drop in harmonic yield for the photoexcited state, the material retains monoclinic symmetry and has not yet transitioned to the rutile. In the future, it should be possible to significantly improve the signal-to-noise ratio of these measurements, using mid-IR optical parametric chirped-pulse amplification (OPCPA) running at kHz repetition rates[44] to further investigate the observed modulations.

In conclusion, we have demonstrated the first instance of tr-HHG spectroscopy from a strongly correlated material. We have shown that tr-HHG is capable of discriminating and tracking the real-time evolution of the distinct phase transitions and associated coherent phonon dynamics in $VO_2$, including observations of the hidden metastable $\mathcal{M}$ phase that has no equilibrium analog. This technique provides complementary information to UED and optical measurements of similar nature. As tr-HHG spectroscopy is highly sensitive to the band structure of a material, it can be used to probe phase transitions in materials as their band structures change. For example, we observe that although the metastable $\mathcal{M}$ phase in $VO_2$ presents with metallic optical properties, the band structure in this phase allows for a high HHG yield that is directly correlated to the $\mathcal{M}$ phase fraction. Due to the sensitivity of tr-HHG and the capability of probing structural dynamics, we expect that tr-HHG spectroscopy can be extended to other strongly correlated materials and other solids, to detect some of the promising features listed in [24] and to gain information about the dynamics as the systems evolve through different electronic states. With a simple experimental setup to implement, the tr-HHG spectroscopy technique developed in this work opens the path to the study of how materials evolve and transform to exotic phases under such different conditions as high pressure, high temperature, and photoexcitation. Moreover, with this setup, we can vary the angle of polarization or incidence, paving the way to retrieve parallel information between tr-HHG and other angle resolved spectroscopy measurements for the study of 2D materials.

**Acknowledgments:** We thank Antoine Laramée for his technical contributions. We thank Misha Ivanov, Andrew Bruhács and Luke Govia for their scientific discussions. **Funding:** We acknowledge funding from NSERC, FRQNT, MESI, and CFI-MSI. V.G. was supported by the Air Force Office of Science Research under MURI Award No. FA9550-16-1-0013. **Author contributions:** M.R.B and F.L conceived the experiments. M.R.B. performed the experiments with assistance from E.H., V.G., A.L., H.I., and P.L. J.C., N.É., and M.C., fabricated and characterized the samples. M.R.B. analyzed the results with support from H.I. and F.L. M.R.B. wrote the manuscript and supplementary information with significant contributions from V.G., A.L., H.I., M.R.O., B.J.S., and F.L. All authors contributed to revising and editing the manuscript.

**Methods:**

**Experimental Setup.**

These experiments were performed on the 50 Hz beamline at the Advanced Laser Light Source (ALLS, Varennes, QC, Canada). A high energy optical parametric amplifier (HE-OPA) was used to generate the mid-infrared (MIR) source used to drive high harmonic generation (HHG) in $VO_2$. This OPA line is pumped by 1.1 mJ of 800 nm, Ti:Sapphire laser at a 50 Hz repetition rate, producing 150 µJ of 1.7 µm pulses using a commercial two-stage TOPAS system (Light Conversion). The output is further amplified in a 20x20x1 mm, Type-I BBO crystal ($\theta$ = 19.9°) using 20 mJ of 800 nm light. From this, 7 mJ of signal (1.5 µm) and idler (1.7 µm) is produced that is then used to drive the difference frequency generation (DFG) stage. A 10 µm MIR source is obtained by DFG in a 100 µm thick GaSe crystal at 30° to produce ~20 µJ of pulse energy. The MIR was characterized using a Spectral Products grating-based slit monochromator, with a liquid-nitrogen-cooled HgCdTe detector, with a central wavelength found to be 10 µm, and with a pulse duration of 80 fs, measured via frequency resolved optical switching[45]. The OPA line can also be tuned such that the DFG generates 7 µm with similar pulse duration and energy to the 10 µm.

A surface reflection from a thin wedge is taken from the signal arm of the DFG setup to pump the phase transitions in $VO_2$ at a wavelength of 1.5 µm. The fluence of this pump is modulated using a half-waveplate, Ge polarizer energy throttle and slightly focused using a 150 mm lens to achieve a fluence on the order of tens of mJ cm$^{-2}$. The delay between the 1.5 µm pump and the MIR driver is varied using a Thorlabs 2825B dc stepper motor actuator. The pulse duration of the pump was characterized using a home-built second harmonic generation frequency resolved optical gating (SHG-FROG) setup and found to be 50 fs full width at half maximum (fwhm).

13 µJ of MIR is focused onto the $VO_2$ sample with an off-axis parabola (OAP) with a reflected focal length of 40 mm. This gives us an intensity of ~$1.7 \times 10^{12}$ W cm$^{-2}$, and a Keldysh parameter much less than one, thus we are in the tunneling regime[46,47]. The HHG from the $VO_2$ driven by 10 µm are collected using an Ocean Optics NIR256 spectrometer starting from the 5$^{th}$ harmonic. Tr-HHG measurements cannot be performed using the 7$^{th}$ harmonic as it is too close to the pump wavelength to be spectrally separated. The 3$^{rd}$ harmonic driven by 7 µm is collected using a Thorlabs DET10D2 biased InGaAs detector with a long pass filter at 1700 nm. Once again, the tr-HHG measurements from the 5$^{th}$ harmonic cannot be separated from the pump wavelength. The HHG spectra were taken using the spectrometer and the amplitude of the 5$^{th}$ harmonic yield was measured using the photodiode. Unless otherwise specified, the experiments were performed in ambient conditions at a room temperature of 20 °C. HHG measurements were taken in both transmission and in the backward direction. Complementary pump-probe transmissivity measurements were performed using the 1.7 µm idler pulse as the probe.

**Sample Preparation.**

The epitaxial $VO_2$ samples were deposited onto a 532 µm *r*-cut sapphire substrate via reactive pulsed laser deposition and grown to a thickness of 100 nm at the Laboratory of Micro and Nanofabrication (LMN, Varennes, QC, Canada). The measured band gap ($E_g$) of the sample is found to be $E_g = 0.683 \pm 0.002$ eV (1815 nm). Details of the deposition process and characterization of the sample can be found in Bionta *et al.*[35].

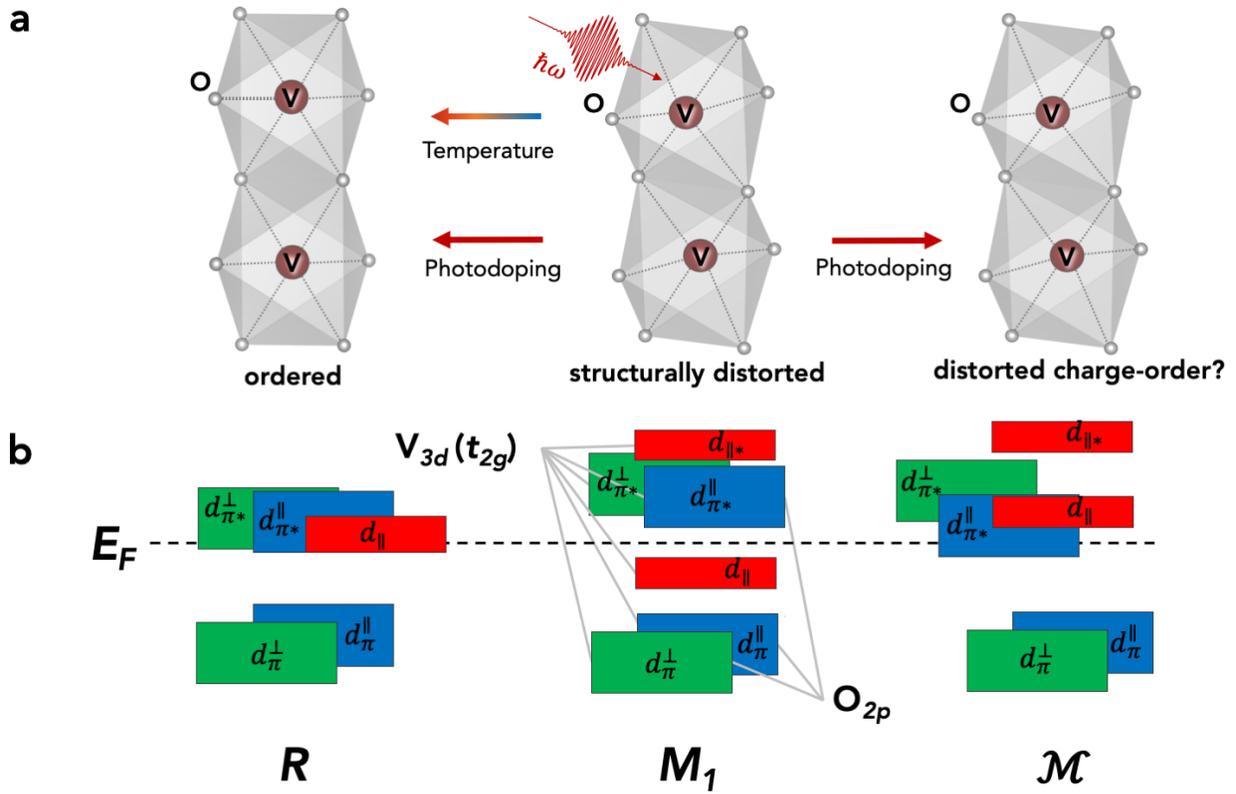

**Fig. 1: Atomic and band structure of VO$_2$ IMT. a** Atomic arrangement of VO$_2$ during IMT phase transitions. In the photoinduced phase transition, the $M_1 \rightarrow R$ transition occurs at high fluence, while the $M_1 \rightarrow \mathcal{M}$ occurs at low to moderate fluences. **b** Energy diagram of the band structure of VO$_2$ in the $M_1$, $R$, and $\mathcal{M}$ phases.

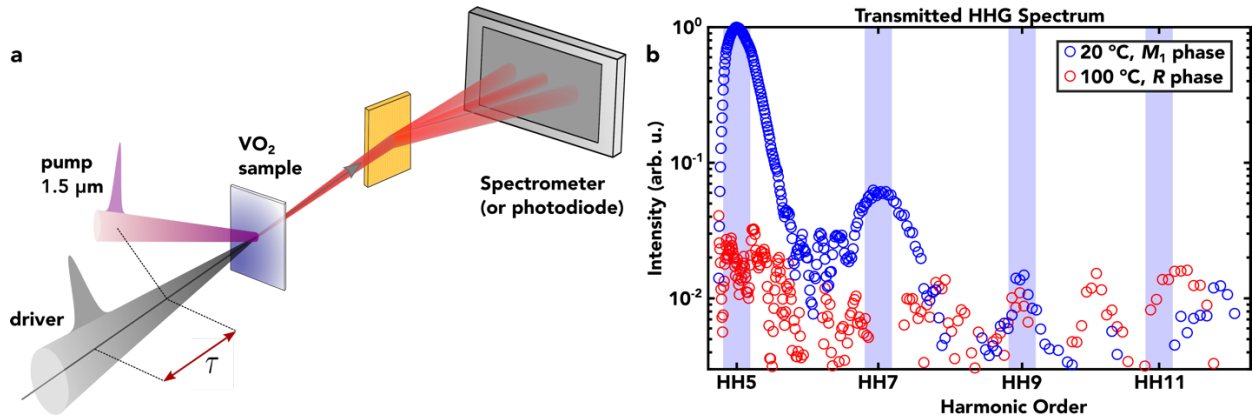

**Fig. 2: Experimental geometry. a** An 80 fs, MIR driver pulse is used to generate harmonics from a 100 nm thick epitaxial VO$_2$ sample. The transmitted harmonic spectrum is then collected and recorded using a spectrometer or photodiode. A 50 fs, 1.5 μm pump at variable delay, τ, from the driver is used to photoexcite the IMT. The fluence of the pump can be modulated using a half-waveplate, polarizer energy throttle. **b** The unpumped high harmonic (HH) spectrum driven by 10 μm in the $M_1$ phase at room temperature of 20 °C (blue circles) and $R$ phase at 100 °C (red circles). Blue rectangles indicate location of expected harmonics.

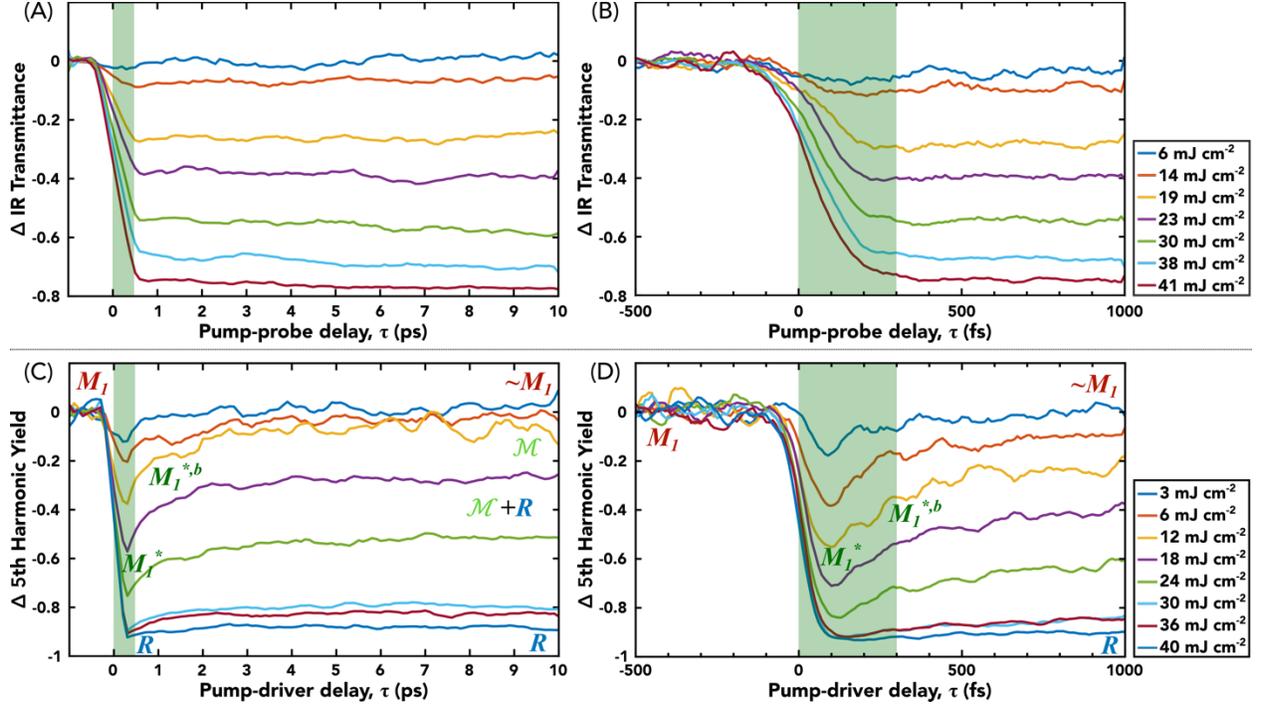

**Fig. 3: Change in optical IR transmittance (top) and fifth harmonic yield (bottom).** The time dependent change in optical IR transmittance (top) or the fifth harmonic yield (bottom) driven by 10 μm following the photoexcited IMT in $VO_2$. Negative delays indicate the harmonic generating driver pulse arrives before photoexcitation. The various phases of $VO_2$ are annotated on the curves. No revival of the IR transmittance is observed for long time-scales in **a** or short, in **b**. **c** Long-term recovery and revival of the harmonics as the $VO_2$ transitions to the $\mathcal{M}$ phase. The amplitude of $\mathcal{M}$ recovery in the HHG yield is plotted as a function of pump fluence in Fig. 4b. The dashed rectangle is expanded upon with higher time-resolution in **d**. The shaded green rectangles show the fast dynamics for the $M_1^* \to M_1^{*,b}$ transition. In both **c** and **d** the suppression of the harmonics in the $R$ phase can be seen.

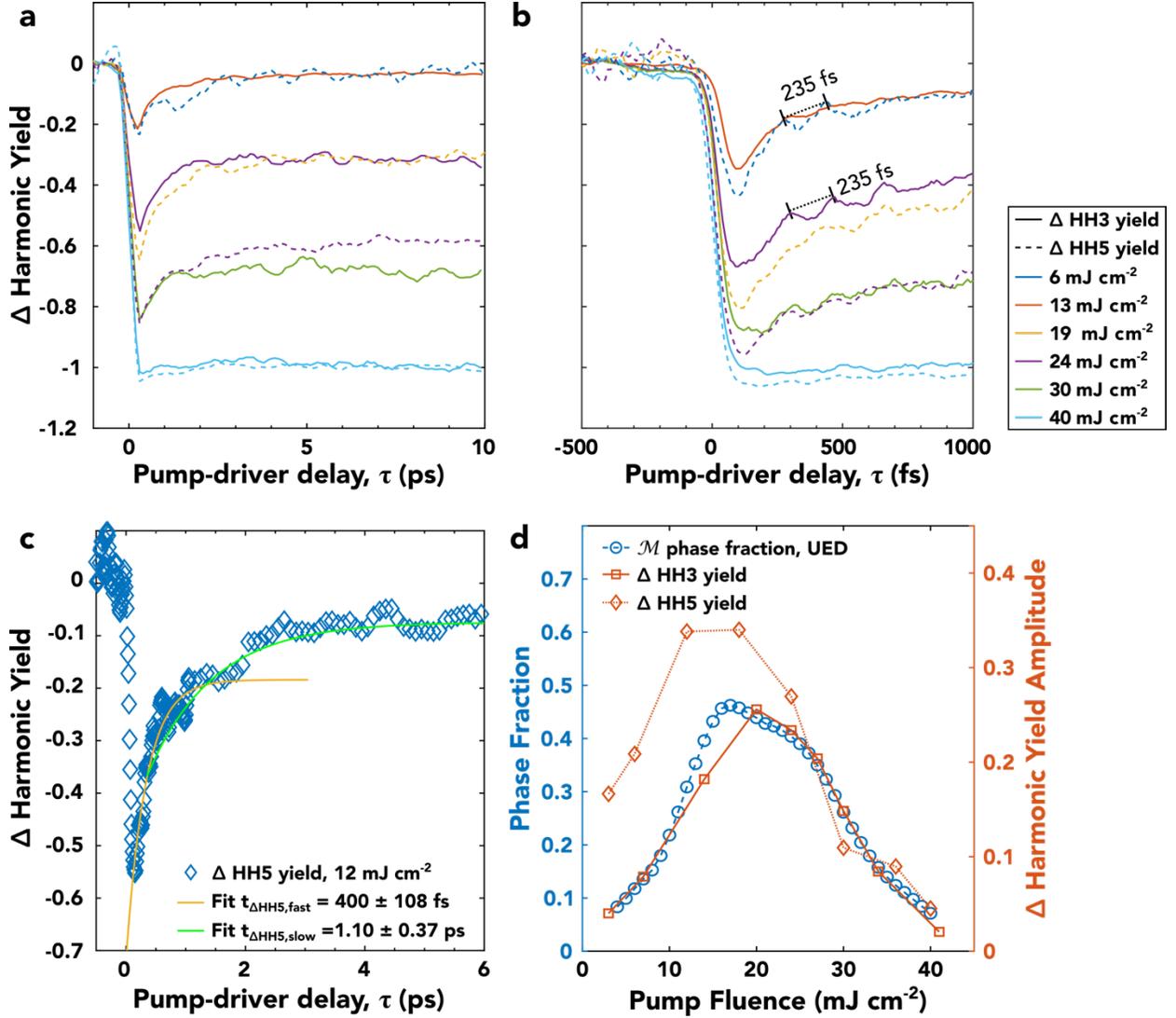

**Fig. 4: HHG and UED comparisons. a** & **b** Comparison of the 5$^{th}$ harmonic (solid lines, driven by 10 μm) and 3$^{rd}$ harmonic (dashed lines, driven by 7 μm) dynamics for varying pump fluences. Similar dynamics are observed regardless of harmonic probed. A 235 fs oscillation is indicated in **b** corresponding to a 4.4 THz phonon mode occurring in $M_1$ phase VO$_2$. **c** Bi-exponential fit of a moderately pumped HHG yields where both the $M_1^* \rightarrow M_1^{*,b}$ and $M_1^{*,b} \rightarrow \mathcal{M}$ transitions are occurring. **d** Extracted phase fractions for the $\mathcal{M}$ phase from UED results [32,33] and the corresponding change in amplitude for the HHG yield.

# Supplementary Information for

# Tracking ultrafast solid-state dynamics using high harmonic spectroscopy

**Authors:** Mina R. Bionta[1, †,*], Elissa Haddad[1], Adrien Leblanc[1], Vincent Gruson[1,2], Philippe Lassonde[1], Heide Ibrahim[1], Jérémie Chaillou[1], Nicolas Émond[1], Martin R. Otto[3], Bradley J. Siwick[3], Mohamed Chaker[1], and François Légaré[1,*]

**Affiliations:**

[1]Centre Énergie Matériaux et Télécommunications, Institut National de la Recherche Scientifique, 1650 Boulevard Lionel-Boulet, Varennes, Québec, J3X 1S2, Canada.

[2]Department of Physics, The Ohio State University, 191 West Woodruff Avenue, Columbus, Ohio, 43210, USA.

[3]Department of Physics and Department of Chemistry, Center for the Physics of Materials, McGill University, 801 Sherbrooke Street W, Montreal, Québec, H3A 2K6, Canada.

[†]Present address: Research Laboratory of Electronics, Massachusetts Institute of Technology, 77 Massachusetts Avenue, Cambridge, Massachusetts, 02139, USA.

[*]Correspondence to: mbionta@mit.edu; legare@emt.inrs.ca

S1. *R* phase measurements

For comparison to the photoexcited phase transition with high fluence to the *R* phase, measurements were performed at a temperature of 100 °C. This temperature is well above the critical temperature for the IMT in $VO_2$ of ~343 K (~70 °C) so the sample is sure to be in the *R* phase. The generated high harmonic spectra in the forward and backward[48] direction are compared for the $M_1$ vs *R* phases in Fig. S1.

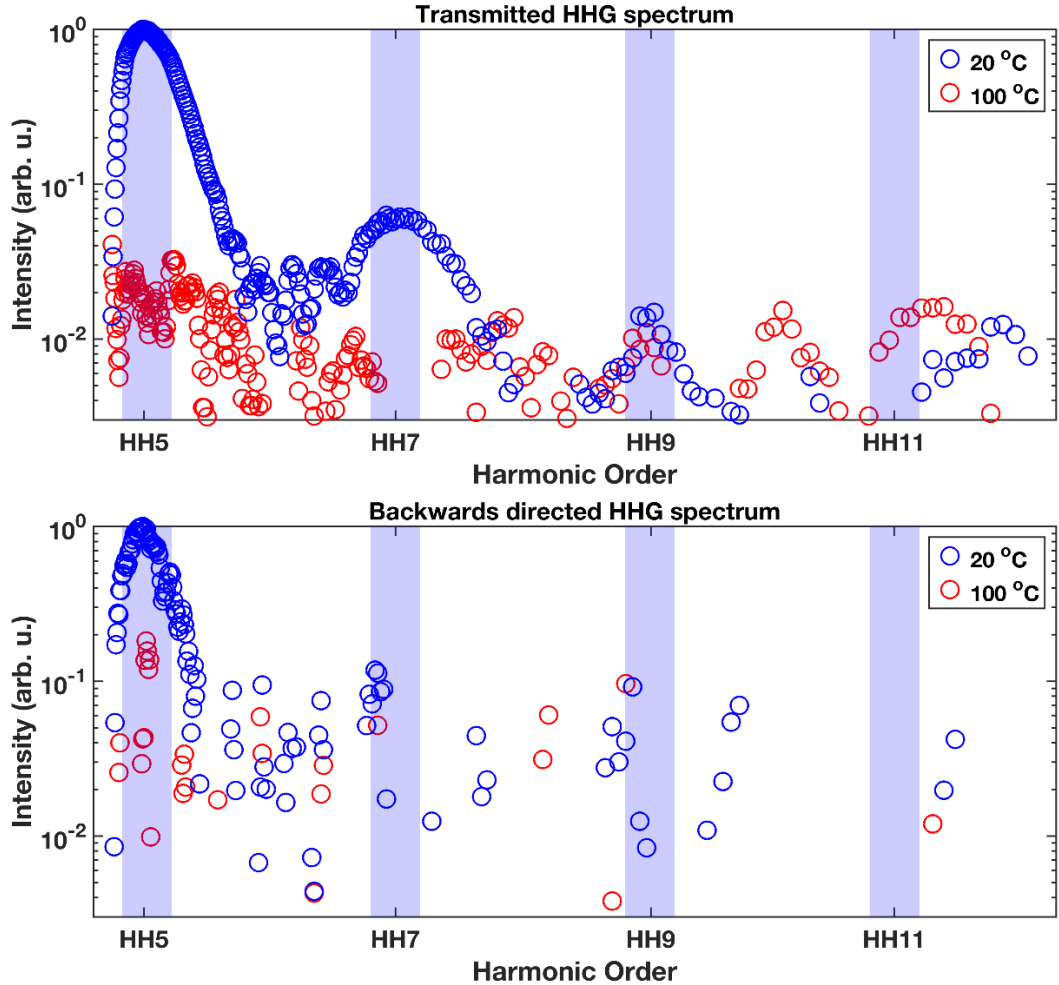

**Fig. S1: $M_1$ vs *R* phase harmonic generation.** High harmonic spectra in the forward (top) and backward (bottom) direction for the $M_1$ phase (blue) at 20 °C and *R* phase (red) at 100 °C. Blue shaded rectangle shows location of expected harmonics. HH: High Harmonic.

S2. Fitting
From the UED and time resolved (tr-) HHG results, timescales for all three phase transitions can be found.

*S2a. UED results from Morrison et al[32]:*
Time constants for the $M_1 \to R$ and $M_1 \to \mathcal{M}$ phase transitions were found with the UED results of Morrison *et al.*[32]. We compared the time constant found for the $M_1 \to \mathcal{M}$ phase transition with that obtained from the tr-HHG.

**$M_1 \to \mathcal{M}$**: The $M_1 \to \mathcal{M}$ time constant was found by fitting the UED results of Morrison *et al.*[32]. They fit a time constant to the changing (200) and (220) peaks that are present for both the monoclinic and rutile phases in $VO_2$ (Fig. S3B, red circles). If an $M_1 \to R$ transition is initiated via photoexcitation, a ~300 fs time constant would be observed for the $(30\bar{2})$ peak. However, a slower process is observed, even at lower pump fluences. This is because the atomic form factor term in the scattering intensity increases due to an electronic reorganization of the electrostatic potential since the diffracted intensity is sensitive to the valence charge distribution. This is attributed to the formation of the $\mathcal{M}$ state. Since the (200) and (220) are relatively low index peaks with a low scattering vector, they are sensitive to long range order in the sample. A time constant for the $M_1 \to \mathcal{M}$ is found to be $t_{slow} = 1.6 \pm 0.2$ ps (Fig. S3B, purple dashed line) when pumped with 20 mJ cm$^{-2}$ at 800 nm.

*S2b. HHG results from this work:*
The two timescales present in the change in harmonic yield were each fitted to a separate equation of the form $S = A_0 - A_1 e^{-(t-t_0)/t_c}$, where $t_c$ is the time constant. This is consistent with the theory analysis of He and Millis[34] who show that the $M_1 \to \mathcal{M}$ phase transition has two components: a fast thermalization transition ($M_1^* \to M_1^{*,b}$), followed by the long relaxation to the $\mathcal{M}$ phase ($M_1^{*,b} \to \mathcal{M}$). These two transitions are probed using the tr-HHG yield.

The time constants for each transition are found as followed:

**$M_1^* \to M_1^{*,b}$**: The $M_1^* \to M_1^{*,b}$ time constant is found when the system is pumped by a low fluence of 3 mJ cm$^{-2}$ (@ 1.5 μm). At this fluence, there is insufficient energy to initiate the phase transition and the system returns to the $M_1$ state after photoexcitation. We fit these features to the 3$^{rd}$ and 5$^{th}$ tr-HHG curves to give us $t_{\Delta HH3,fast} = 300 \pm 24$ fs and $t_{\Delta HH5,fast} = 245 \pm 68$ fs (Fig. S2) respectively, which is in good agreement of the ~100s fs found by He and Millis[34].

**$M_1^{*,b} \to \mathcal{M}$**: Looking at an IMT pumped with moderate fluence (12-20 mJ cm$^{-2}$ @ 1.5 μm), we are able to clearly see the $M_1^{*,b} \to \mathcal{M}$ timescales from our datasets (Fig. S3A green curve) and are in good agreement with the UED results of ref. [32] (Fig. S3B). We can extract a slow timescale for the $M_1^{*,b} \to \mathcal{M}$ transition of $t_{\Delta HH3,slow} = 1.45 \pm 0.34$ ps and $t_{\Delta HH5,slow} = 1.10 \pm 0.37$ ps for the tr-HHG yields of the 3$^{rd}$ and 5$^{th}$ harmonics.

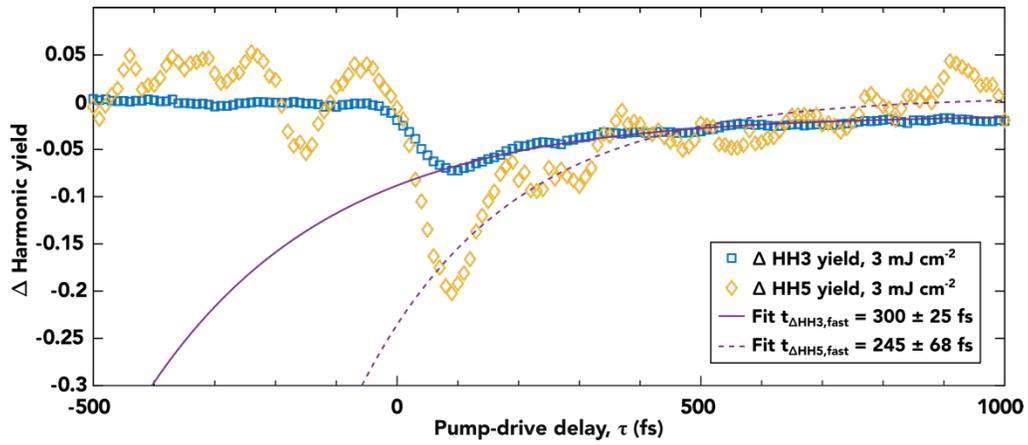

**Fig. S2: Fast timescale** Fast time dynamics for the $M_1^* \to M_1^{*,b}$ transition of $t_{fast} = 300 \pm 25$ fs and $245 \pm 68$ fs retrieved from the 3rd and 5th tr-HHG measurements respectively. Each was pumped with only 3 mJ cm$^{-2}$ where this is the only transition present.

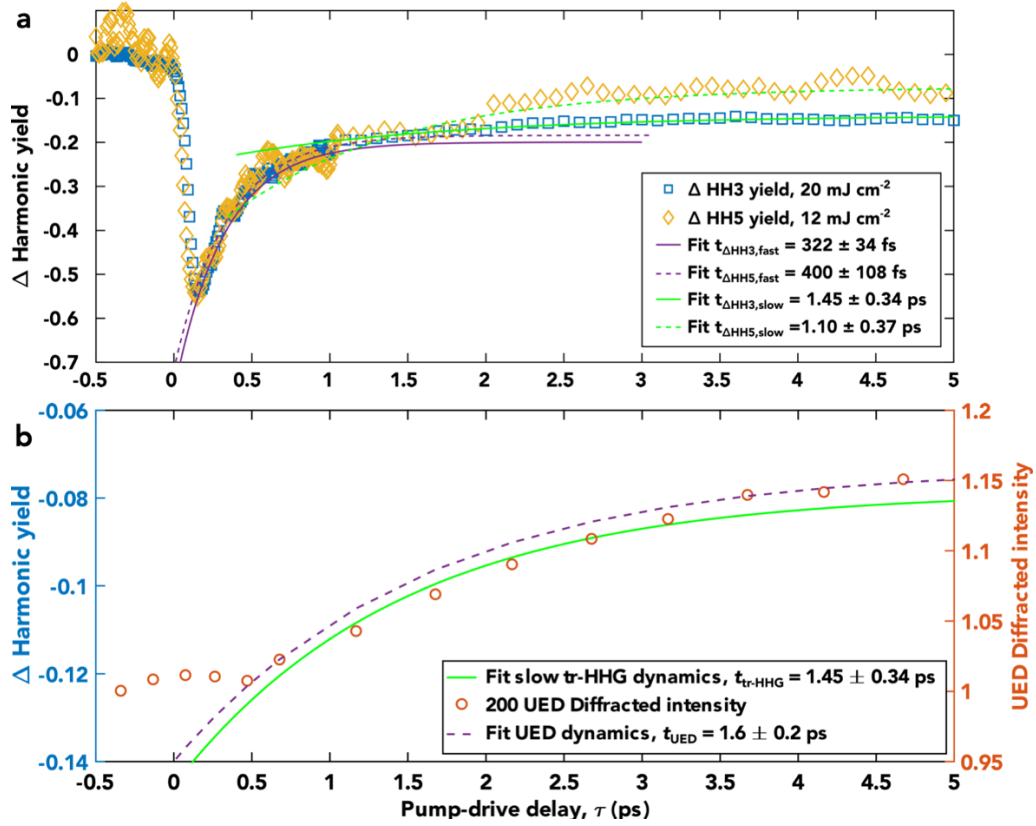

**Fig. S3: Two timescales. a** Initiating the IMT with a moderate pump fluence of 12-20 mJ cm$^{-2}$ (@ 1.5 μm) presents both the $M_1^* \to M_1^{*,b}$ and $M_1^{*,b} \to \mathcal{M}$ timescales in the changing harmonic yield for both the 3$^{rd}$ and 5$^{th}$ tr-HHG measurements. Fast dynamics of $t_{fast}$ = 322 ± 34 fs and 400 ± 108 fs (purple curves), and slow dynamics of $t_{slow}$ = 1.45 ± 0.34 ps and 1.10 ± 0.37 ps (green curves) are retrieved from the 3$^{rd}$ and 5$^{th}$ harmonic curves respectively. **b** The slow timescale retrieved from 3$^{rd}$ harmonic tr-HHG (green curve) is in good agreement with the $t_{UED}$ = 1.6 ± 0.2 ps results from UED (data red circles, fit purple dashed curve).

S3. Backward direction harmonic measurements

While the measurements presented in the main text were performed in transmission, the same tr-HHG measurements for the 5$^{th}$ harmonic of 10 μm were realized in the backward direction[48], demonstrating the same behavior with a recovery of the harmonic yield for low and intermediate pump fluences (Fig. S4).

Backward emitted harmonics are generated on the front surface of the material without propagation through the sample, thus the evolution of the harmonic yield detected originates from variations in the material state. When modulating the pump fluence initiating the IMT, we see the same temporal yield dependency in the backward emitted harmonic generation as in the forward direction. This indicates that the changing states of $VO_2$ actually affects the production of the harmonics, not simply changing the transmission of the harmonics through the sample.

As we are able to detect harmonics that are generated in reflection, we can establish that the harmonics are generated on the surface of the sample. Although the signal-to-noise ratio is lower in reflection as there is less signal to collect, the recovery in harmonic yield to the $\mathcal{M}$ state is clearly present alongside the two timescales for transition from the $M_1^* \to M_1^{*,b}$ and $M_1^{*,b} \to \mathcal{M}$ states.

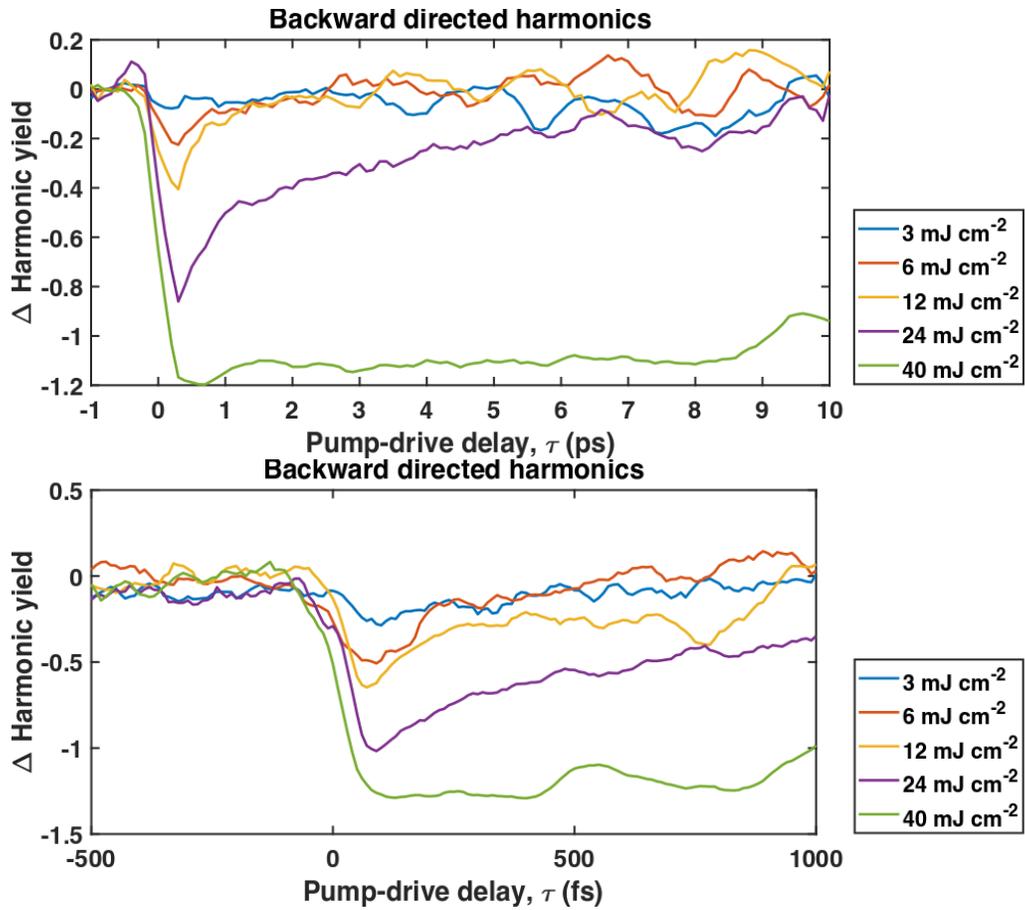

**Fig. S4: Harmonic yield in the backwards direction.** These curves show the change in harmonic yield for harmonics generated in the backwards direction. We can still see the same double timescale that is observed in transmission.

**Supplementary References:**